\documentclass{book}    
\usepackage{piers}  
\begin{document}

\title{Influence of field potential on the speed of light}
\maketitle

\author      {Zi-Hua Weng}
\affiliation {Xiamen University}
\address     {}
\city        {Xiamen}
\postalcode  {}
\country     {China}
\phone       {345566}    
\fax         {233445}    
\email       {xmuwzh@xmu.edu.cn}  
\misc        { }  
\nomakeauthor

\begin{authors}

{\bf Zi-Hua Weng}\\
\medskip
School of Physics and Mechanical \& Electrical Engineering,
\\Xiamen University, Xiamen 361005, China\\

\end{authors}

\begin{paper}

\begin{piersabstract}
The paper discusses the affection of the scalar potential on the
speed of light in the electromagnetic field, by means of the
characteristics of octonion. In the octonion space, the radius
vector is combined with the integral of field potentials to become
one new radius vector. When the field potentials can not be
neglected, the new radius vector will cause the prediction to
departure slightly from the theoretical value of the speed of light.
The results explain why the speed of light varies in diversiform
optical waveguide. And there exist negative refractive indexes due
to different scalar potentials in the gravitational field and
electromagnetic field.
\end{piersabstract}


\psection{Introduction}

The invariable speed of light is being doubted all the time. And
this question remains as puzzling as ever. But the existing theories
do not clarify why the speed of light has to keep unchanged, and
then do not offer reasonable explain for this empirical fact. The
paper attempts to reason out why the speed of light keeps the same
in most cases, even in the electromagnetic field.

The invariable speed of light has not been validated in strong
electromagnetic field, although the speed of light is variable in
optical waveguide materials. Some experiments for the variable speed
of light have been performed by L. V. Hau \cite{hau}, M. L.
Povinelli \cite{povinelli}, and J. W. Moffat \cite{moffat}, etc. But
all of these verifications are not dealt with the electromagnetic
field potential, and have been validated in neither strong
electromagnetic field nor gravitational field. So this puzzle of
invariable speed of light remains unclear and has not satisfied
results.

The algebra of quaternions \cite{adler} was first used by J. C.
Maxwell to describe the electromagnetic field. The octonions
\cite{cayley} can be used to demonstrate the electromagnetic field
and gravitational field simultaneously \cite{weng}. In the octonion
space, the speed of light will be varied with the electromagnetic
field potential as well as gravitational field potential. And there
exist negative refractive indexes in optical waveguide materials,
when the electromagnetic field potential is switched from positive
to negative. This inference is coincided with that in the negative
index materials \cite{soukoulis}.

\psection{Octonion transformation}

In the octonion space, the basis vector $\mathbb{E}$ consists of the
quaternion basis vectors $\mathbb{E}_g$ and $\mathbb{E}_e$ . The
basis vector $\mathbb{E}_g = (1, \emph{\textbf{i}}_1,
\emph{\textbf{i}}_2, \emph{\textbf{i}}_3$) is the basis vector of
the quaternion space for the gravitational field, and $\mathbb{E}_e
= (\emph{\textbf{I}}_0, \emph{\textbf{I}}_1, \emph{\textbf{I}}_2,
\emph{\textbf{I}}_3$) for the electromagnetic field. And that the
basis vector $\mathbb{E}_e$ is independent of the $\mathbb{E}_g$,
with $\mathbb{E}_e$ = $\mathbb{E}_g \circ \emph{\textbf{I}}_0$ .
\begin{eqnarray}
\mathbb{E} = (1, \emph{\textbf{i}}_1, \emph{\textbf{i}}_2,
\emph{\textbf{i}}_3, \emph{\textbf{I}}_0, \emph{\textbf{I}}_1,
\emph{\textbf{I}}_2, \emph{\textbf{I}}_3)
\nonumber
\end{eqnarray}

The octonion physical quantity $\mathbb{D} (d_0, d_1, d_2, d_3, D_0,
D_1, D_2, D_3 )$ is defined as follows.
\begin{eqnarray}
\mathbb{D} = d_0 + \Sigma (d_j \emph{\textbf{i}}_j) + \Sigma (D_i
\emph{\textbf{I}}_i)
\end{eqnarray}
where, $d_i$ and $D_i$ are all real; $i = 0, 1, 2, 3$; $j , k = 1,
2, 3$.

When the octonion coordinate system is transformed into the other,
the physical quantity $\mathbb{D}$ will be transformed into the
octonion $\mathbb{D}' (d'_0 , d'_1 , d'_2 , d'_3 , D'_0 , D'_1 ,
D'_2 , D'_3 )$.
\begin{eqnarray}
\mathbb{D}' = \mathbb{K}^* \circ \mathbb{D} \circ \mathbb{K}
\end{eqnarray}
where, $\mathbb{K}$ is the octonion, and $\mathbb{K}^* \circ
\mathbb{K} = 1$; $*$ denotes the conjugate of octonion; $\circ$ is
the octonion multiplication.

When the spatial coordinates $d_1, d_2, d_3, D_0, D_1, D_2, D_3$
take part in the rotation, the octonion $\mathbb{D}$ satisfies the
following relation.
\begin{eqnarray}
d_0 = d'_0
\end{eqnarray}

In the above equation, the scalar part $d_0$ is preserved during the
octonion spatial coordinates are transforming. Some invariants of
electromagnetic field will be obtained from the characteristics of
the octonion physical quantity.

\begin{table}[h]
\caption{The octonion multiplication table.} \label{tab:table1}
\centering
\begin{tabular}{ccccccccc}
\hline \hline $ $ & $1$ & $\emph{\textbf{i}}_1$  &
$\emph{\textbf{i}}_2$ & $\emph{\textbf{i}}_3$  &
$\emph{\textbf{I}}_0$  & $\emph{\textbf{I}}_1$
& $\emph{\textbf{I}}_2$  & $\emph{\textbf{I}}_3$  \\
\hline $1$ & $1$ & $\emph{\textbf{i}}_1$  & $\emph{\textbf{i}}_2$ &
$\emph{\textbf{i}}_3$  & $\emph{\textbf{I}}_0$  &
$\emph{\textbf{I}}_1$
& $\emph{\textbf{I}}_2$  & $\emph{\textbf{I}}_3$  \\
$\emph{\textbf{i}}_1$ & $\emph{\textbf{i}}_1$ & $-1$ &
$\emph{\textbf{i}}_3$  & $-\emph{\textbf{i}}_2$ &
$\emph{\textbf{I}}_1$
& $-\emph{\textbf{I}}_0$ & $-\emph{\textbf{I}}_3$ & $\emph{\textbf{I}}_2$  \\
$\emph{\textbf{i}}_2$ & $\emph{\textbf{i}}_2$ &
$-\emph{\textbf{i}}_3$ & $-1$ & $\emph{\textbf{i}}_1$  &
$\emph{\textbf{I}}_2$  & $\emph{\textbf{I}}_3$
& $-\emph{\textbf{I}}_0$ & $-\emph{\textbf{I}}_1$ \\
$\emph{\textbf{i}}_3$ & $\emph{\textbf{i}}_3$ &
$\emph{\textbf{i}}_2$ & $-\emph{\textbf{i}}_1$ & $-1$ &
$\emph{\textbf{I}}_3$  & $-\emph{\textbf{I}}_2$
& $\emph{\textbf{I}}_1$  & $-\emph{\textbf{I}}_0$ \\
\hline $\emph{\textbf{I}}_0$ & $\emph{\textbf{I}}_0$ &
$-\emph{\textbf{I}}_1$ & $-\emph{\textbf{I}}_2$ &
$-\emph{\textbf{I}}_3$ & $-1$ & $\emph{\textbf{i}}_1$
& $\emph{\textbf{i}}_2$  & $\emph{\textbf{i}}_3$  \\
$\emph{\textbf{I}}_1$ & $\emph{\textbf{I}}_1$ &
$\emph{\textbf{I}}_0$ & $-\emph{\textbf{I}}_3$ &
$\emph{\textbf{I}}_2$  & $-\emph{\textbf{i}}_1$
& $-1$ & $-\emph{\textbf{i}}_3$ & $\emph{\textbf{i}}_2$  \\
$\emph{\textbf{I}}_2$ & $\emph{\textbf{I}}_2$ &
$\emph{\textbf{I}}_3$ & $\emph{\textbf{I}}_0$  &
$-\emph{\textbf{I}}_1$ & $-\emph{\textbf{i}}_2$
& $\emph{\textbf{i}}_3$  & $-1$ & $-\emph{\textbf{i}}_1$ \\
$\emph{\textbf{I}}_3$ & $\emph{\textbf{I}}_3$ &
$-\emph{\textbf{I}}_2$ & $\emph{\textbf{I}}_1$  &
$\emph{\textbf{I}}_0$  & $-\emph{\textbf{i}}_3$
& $-\emph{\textbf{i}}_2$ & $\emph{\textbf{i}}_1$  & $-1$ \\
\hline
\end{tabular}
\end{table}

\psection{Speed of gravitational intermediate boson}

In the case for coexistence of the electromagnetic field and the
gravitational field, the algebra of octonions can be used to
describe the property of electromagnetic field and gravitational
field.

According to the viewpoint of field theories, each fundamental
interaction is mediated by the exchange of its intermediate bosons
between particles. The gravitational interaction is mediated by the
exchange of gravitational intermediate bosons between masses.
Meanwhile the electromagnetic interaction is mediated by the
exchange of electromagnetic intermediate bosons between charges. And
that the gravitational intermediate boson and the electromagnetic
intermediate boson can be combined together to become the photon.
The latter can be interacted with either gravitational field or
electromagnetic field.

With the feature of octonions, we find that the gravitational field
potential has an influence on the speed of gravitational
intermediate boson in the gravitational field. It means that the
speed of gravitational intermediate boson is variable in the case
for coexistence of the electromagnetic field and gravitational
field, under the octonion coordinate transformation.

\psubsection{Radius vector}

In the octonion space for gravitational field and electromagnetic
field, the octonion radius vector $\mathbb{R} = \Sigma (r_i
\emph{\textbf{i}}_i) + \Sigma (R_i \emph{\textbf{I}}_i)$. And that
it can be combined with the octonion $\mathbb{X} = \Sigma (x_i
\emph{\textbf{i}}_i) + \Sigma (X_i \emph{\textbf{I}}_i)$ to become
one new radius vector $\bar{\mathbb{R}}= \Sigma (\bar{r}_i
\emph{\textbf{i}}_i) + \Sigma (\bar{R}_i \emph{\textbf{I}}_i)$. The
$\mathbb{X}$ is the integral of field potentials.
\begin{eqnarray}
\mathbb{\bar{R}} = \mathbb{R} + k_{rx} \mathbb{X}
\end{eqnarray}
where, $\emph{\textbf{i}}_0 = 1$; $\bar{r}_i = r_i + k_{rx} x_i$;
$\bar{R}_i = R_i + k_{eg} k_{rx} X_i$; $r_0 = v_0 t$; $R_0 = V_0 T$;
$k_{rx} = 1$.  $t$ denotes the time, $T$ is a time-like quantity.
$v_0$ is the speed of gravitational intermediate boson; $V_0$ is the
speed of electromagnetic intermediate boson. $\mu_e$ and $\mu_g$ are
the coefficients for the electromagnetic field and gravitational
field respectively. $k_{eg}$ is a coefficient, and $k_{eg}^2 = \mu_g
/ \mu_e$ .

In other words, the $\mathbb{\bar{R}}$ can be considered as the
radius vector in the octonion space, with the basis vector $(1,
\emph{\textbf{i}}_1, \emph{\textbf{i}}_2, \emph{\textbf{i}}_3,
\emph{\textbf{I}}_0, \emph{\textbf{I}}_1, \emph{\textbf{I}}_2,
\emph{\textbf{I}}_3)$. When the octonion coordinate system is
rotated, we obtain the radius vector $\mathbb{\bar{R}}' (\bar{r}'_0,
\bar{r}'_1, \bar{r}'_2, \bar{r}'_3, \bar{R}'_0, \bar{R}'_1,
\bar{R}'_2, \bar{R}'_3 )$. From Eqs.(3) and (4), we have
\begin{eqnarray}
\bar{r}_0 = \bar{r}'_0~.
\end{eqnarray}

The above states that the scalar $\bar{r}_0$ remains unchanged when
the coordinate system rotates in the octonion space. And that there
may exist the special case of the $x_i \neq 0$ when $r_i = R_i = 0$.

\psubsection{Velocity}

The velocity $\mathbb{V} = \Sigma (v_i \emph{\textbf{i}}_i) + \Sigma
(V_i \emph{\textbf{I}}_i)$ and the field potential $\mathbb{A} =
\Sigma (a_i \emph{\textbf{i}}_i) + k_{eg} \Sigma (A_i
\emph{\textbf{I}}_i)$ can be combined together to become one new
velocity $\mathbb{\bar{V}} = \Sigma (\bar{v}_i \emph{\textbf{i}}_i)
+ \Sigma (\bar{V}_i \emph{\textbf{I}}_i)$ in the octonion space.
\begin{eqnarray}
\mathbb{\bar{V}} = \mathbb{V} + k_{rx} \mathbb{A}
\end{eqnarray}
where, $\bar{v}_i = v_i + k_{rx} a_i$; $\bar{V}_i = V_i + k_{eg}
k_{rx} A_i$; $a_0$ and $A_0$ are the gravitational scalar potential
and electromagnetic scalar potential respectively.

In the above, the field potential $\mathbb{A}$ consists of the
gravitational field potential $\mathbb{A}_g = \Sigma (a_i
\emph{\textbf{i}}_i)$, and the electromagnetic field potential
$\mathbb{A}_e = \Sigma (A_i \emph{\textbf{I}}_i)$.
\begin{eqnarray}
\mathbb{A} = \mathbb{A}_g + k_{eg} \mathbb{A}_e
\end{eqnarray}

When the coordinate system is rotated, we have one velocity
$\mathbb{\bar{V}}' (\bar{v}'_0, \bar{v}'_1, \bar{v}'_2, \bar{v}'_3,
\bar{V}'_0, \bar{V}'_1, \bar{V}'_2, \bar{V}'_3 )$. From Eqs.(3) and
(6), we have the invariant about the speed of gravitational
intermediate boson in the octonion space.
\begin{eqnarray}
\bar{v}_0 = \bar{v}'_0
\end{eqnarray}

The above means that the speed of gravitational intermediate boson,
$v_0$, will be variable, due to the existence of the scalar
potential, $a_0$, of the gravitational field. Obviously, it is not
associated with the field potential of electromagnetic field.

\psection{Speed of electromagnetic intermediate boson}

In the octonion space for the electromagnetic field and
gravitational field, with the property of the algebra of octonions,
we find that the electromagnetic field potential has an effect on
the speed of electromagnetic intermediate boson in the
electromagnetic field. It states that the speed of electromagnetic
intermediate boson is variable in the case for coexistence of the
electromagnetic field and gravitational field, under the octonion
coordinate transformation.

\psubsection{Radius vector}

In the octonion space, one new octonion quantity $\mathbb{\bar{R}}_q
= \mathbb{\bar{R}} \circ \emph{\textbf{I}}_0^*$ can be defined from
Eq.(4).
\begin{eqnarray}
\mathbb{\bar{R}}_q = \Sigma (\bar{R}_i \emph{\textbf{i}}_i) - \Sigma
(\bar{r}_i \emph{\textbf{I}}_i)
\end{eqnarray}

When the coordinate system is rotated, we have the radius vector
$\mathbb{\bar{R}}'_q ( \bar{R}'_0, \bar{R}'_1, \bar{R}'_2,
\bar{R}'_3, \bar{r}'_0, \bar{r}'_1, \bar{r}'_2, \bar{r}'_3 )$. From
Eqs.(3) and (9), we have
\begin{eqnarray}
\bar{R}_0 = \bar{R}'_0~.
\end{eqnarray}

The above states that the scalar $\bar{R}_0$ remains unchanged when
the coordinate system rotates in the octonion space. And it is easy
to find Eq.(5) and Eq.(10) can not be established simultaneously.

\psubsection{Velocity}

In the octonion space, one new octonion quantity $\mathbb{\bar{V}}_q
= \mathbb{\bar{V}} \circ \emph{\textbf{I}}_0^*$ can be defined from
Eq.(6).
\begin{eqnarray}
\mathbb{\bar{V}}_q = \Sigma (\bar{V}_i \emph{\textbf{i}}_i) - \Sigma
(\bar{v}_i \emph{\textbf{I}}_i)
\end{eqnarray}

When the coordinate system is rotated, we have the velocity
$\mathbb{\bar{V}}' ( \bar{V}'_0, \bar{V}'_1, \bar{V}'_2, \bar{V}'_3,
\bar{v}'_0, \bar{v}'_1, \bar{v}'_2, \bar{v}'_3 )$. From Eqs.(3) and
(11), we have the invariant about the speed of electromagnetic
intermediate boson in the octonion space.
\begin{eqnarray}
\bar{V}_0 = \bar{V}'_0
\end{eqnarray}

The above means that the speed of electromagnetic intermediate
boson, $V_0$, will be variable, due to the existence of the scalar
potential, $A_0$, of the electromagnetic field. Correspondingly, it
is not dealt with the field potential of gravitational field. And
Eq.(8) and Eq.(12) can not be established simultaneously also.

\psection{Speed of light}

In some cases, the electric charge is combined with the mass to
become the electron or proton etc., therefore we have the condition
$ \bar{R}_i \emph{\textbf{I}}_i = \bar{r}_i \emph{\textbf{i}}_i
\circ \emph{\textbf{I}}_0$ and $\bar{V}_i \emph{\textbf{I}}_i =
\bar{v}_i \emph{\textbf{i}}_i \circ \emph{\textbf{I}}_0$ . It means
that the gravitational field as well as the electromagnetic field
has an influence on the movement of the electric charge with the
mass. In other words, those electric charges with the masses take
part in either gravitational interaction or electromagnetic
interaction.

Similarly, the gravitational intermediate boson and the
electromagnetic intermediate boson can be combined together to
become the photon. While, these photons participate not only
gravitational interaction but also electromagnetic interaction. As a
result, the gravitational field potential and electromagnetic field
potential both can impact the speed of light from Eqs.(8) and (12).

In the gravitational theory, the gravitational field potential has
an effect on the speed of light. The inference is similar to the
shift of spectral-line in Einstein's general relativity. In
Maxwell's electromagnetic theory, the electromagnetic field
potential has an influence on the speed of light in the glass etc.
Therefore we have the concept of refractivity in the optics theory.

According to the viewpoint about the affection of field potentials
on the speed of light, there may exist the negative refractive index
due to different field potential. The conclusion may explain why
there exist negative index materials or left-handed materials in a
different way.

\psection{Conclusion}

In the octonion space, the inferences about speed of light depend on
the combinations of physical definitions. By means of definition
combination of radius vector and velocity, the gravitational field
potential as well as electromagnetic field potential are found to
have the influence on the speed of light, in the case for
coexistence of the gravitational field and electromagnetic field.

The speed of light changes with the gravitational field potential as
well as electromagnetic field potential, and has a deviation from
its theoretical value. The light speed variation has a limited
effect on the movement of light, because the variation is quite
small. Therefore the invariable speed of light is believed to be
correct in most cases. However, when there is a very high potential
of electromagnetic field, the light speed variation will become huge
enough to impact the refractive index of materials obviously. There
exist negative refractive indexes in optical waveguide materials,
when the electromagnetic field potential is switched from positive
to negative, or otherwise. This result is coincided with that in the
negative index materials.

It should be noted that the study for influence of field potentials
on the speed of light examined only one simple case with very weak
field potentials in the gravitational field and electromagnetic
field. Despite its preliminary characteristics, this study can
clearly indicate that the field potentials in the gravitational
field and electromagnetic field have an influence on the scalar
invariants. For the future studies, the related investigation will
concentrate on only the predictions of light speed variation due to
the huge field potentials in the gravitational field and
electromagnetic field.

\ack
This project was supported partially by the National Natural
Science Foundation of China under grant number 60677039.

\end{paper}

\end{document}